# PHASE TRANSITIONS AND ANTIFERROELECTRICITY IN BiFeO$_3$ FROM ATOMIC LEVEL SIMULATIONS

M. Graf[1], M. Sepliarsky[1], S. Tinte[2] and M.G. Stachiotti[1]

[1] Instituto de Física Rosario, Universidad Nacional de Rosario, Rosario, Argentina

[2] Instituto de Física del Litoral, Universidad Nacional del Litoral, Santa Fe, Argentina

e-mail: sepliarsky@ifir-conicet.gov.ar

The structural and polar properties of BiFeO$_3$ at finite temperature are investigated using an atomistic shell model fitted to first-principles calculations. Molecular Dynamics simulations show a direct transition from the low-temperature *R3c* ferroelectric phase to the *Pbnm* orthorhombic phase without evidence of any intermediate bridging phase between them. The high-temperature phase is characterized by the presence of two sublattices with opposite polarizations, and it displays the characteristic double-hysteresis loop under the action of an external electric field. The microscopic analysis reveals that the change in the polar direction and the large lattice strains observed during the antiferroelectric-ferroelectric phase transition originate from the interplay between polarization, oxygen octahedron rotations and strain. As a result, the induced ferroelectric phase recovers the symmetry of the low temperature *R3c* phase.

PACs: 77.80.-e, 77.84.-s, 77.80.B

# I. Introduction

The perovskite oxide BiFeO$_3$ (BFO) is attracting much attention due to its unique properties and potential applications. BFO is the most promising magnetoelectric multiferroic material which displays simultaneously ferroelectric (FE) and ferromagnetic properties coupled at room temperature [1–3], and it is expected to have important applications in various fields from spintronics to energy-saving technologies. [4] The interest in BFO is not restricted to its magnetoelectric properties. Its lead-free composition, large switchable polarization and high Curie temperature make it an attractive ferroelectric compound to environmentally friendly devices. [5,6]

A large number of studies were performed in order to elucidate the behavior of BFO as function of the temperature, and much progress was achieved. Nevertheless, the compound has shown a very intricate behavior and there are still many unanswered questions. In particular, we can highlight the behavior at high temperatures, where several aspects remain contradictory or are poorly understood. BFO has a ferroelectric ground state with *R3c* symmetry which remains stable up to T$_C$=1100 K. At this temperature, the system undergoes a first order transition to a paraelectric phase. The nature of the high temperature phase was under debate for a long time, and many structures were proposed based in experiments and in theoretical approaches [3]. It is now accepted that the phase has *Pbnm*-symmetry structure (formally *Pnma*). Nevertheless, the presence of an intermediate bridging region between the rhombohedral and the orthorhombic phases is not completely ruled out [7]. In addition, little is know about the polar properties of the phase(s) above T$_C$. The phase of BFO has always been seen as paraelectric in experiments, even though some signals related to an

AFE order have been suggested. [8–11] The clarification of that region in the phase diagram is, then, of fundamental interest and necessary to understand properties of the single compound [10,12] and its solid solutions [13–15].

A complete experimental characterization of BFO at high temperature is cumbersome due to decomposition and leakage problems, so theoretical simulations are highly desirable. First-principle calculations have contributed greatly to the understanding of the electronic structure and structural instabilities of BFO [9,16–19]. Since these methods are mainly restricted to studying zero-temperature properties, it is necessary to combine them with simple models in order to investigate finite temperature effects in large size systems. Recently, two different theoretical approaches were developed to investigate finite-temperatures properties in BFO [7,20]. Although both methods were able to reproduce the observed low-temperature behavior, they display different results above $T_C$. While the description based in a model Hamiltonian [7] predicts a new family of phases between the *R3c* and the *Pbnm* phase, a direct transition is observed in the simulations performed with a bond valence model [20]. Many factors can contribute to those discrepancies, and it is not clear which approach offers the most reliable predictions. In any case, these results reflect the intrinsic complex structure of the compound, and new studies are necessary in order to clarify especially the behavior of the high-temperature phase. In this paper, we investigate the structural and polar properties of BFO at finite temperature by using an atomistic shell model fitted to first-principles calculations.

## II. MODEL DEVELOPMEMNT

Atomic level simulations with the shell-model approach have been extensively used to study ferroelectric oxides [21–23]. An inherent advantage of this approach is that there is not need to know a priori which dynamical modes are relevant in determining specific material properties. However, the quality of the model is very sensitive to the parameter values, and then a deep validation is necessary. In the shell model, atoms are thought to consist of an ion core coupled to an "electronic" shell in order to include its electronic polarization. In this work, the core-shell coupling is described through an anharmonic spring described as $V(\omega) = 1/2 k_2 \omega^2 + 1/24 k_4 \omega^4$, where $\omega$ is the core–shell displacement. Besides the "intra-atomic" core-shell coupling, the model includes electrostatic interactions among cores and shells of different atoms, and short-range interactions between shells described by a Rydberg potential, $V(r)=(a+Br)\ exp(-r/\rho)$ with a shifted-force correction and a cutoff radius of 7.2 Å. In addition, a Fe-O-Fe three-body bond bending term is included to improve the directional character of that interaction [24,25], which is represented by $V_B(\theta) = ½\ k_B\ (cos(\theta)-cos(\theta_0))^2$ where $\theta_0$ is the equilibrium bond angle. We found that this additional three body term contributes mainly to improve the description of the model under pressure.

The input data to adjust the model parameters were obtained using the local-density approximation (LDA) to the density functional theory as implemented in the *ab-initio* VASP package [26], including the so-called LDA+U correction of Dudarev *et al.* [27] for a better treatment of iron's *3d* electrons. A G-type antiferromagnetic arrangement of the Fe atom spins is assumed. Calculations were made considering a 40-atom supercell, which allows to model the low-symmetry and low-energy structures that the compound presents [9]. The data correspond to values of total energies, forces on atoms and stresses of a total of 40 different configurations. The database includes relevant configurations such as the *R3c* ground state and relaxed structures with *Pbnm* and *Pm3m* symmetry. It also includes configurations with *Imma* symmetry, and different atomic displacements and strained structures. The model parameters were adjusted using a least-square procedure following a similar methodology as in Ref [28]. Since BFO exhibits a large variety of metastable phases in a wide range of energies, a greater weigh was assigned to the configurations with lower energy. The parameters of the resulting potentials are shown in Table I.

A first test on the quality of the developed model was made by relaxing structures with different symmetries to compare with first-principles results. The structural parameters obtained for the *R3c* and *Pbnm* phases are listed in Table 2. First, note that the model is able to reproduce correctly the ground state of BFO yielding as the lowest-energy configuration the rhombohedral *R3c* phase. Its lattice parameter coincides with the *LDA+U* result of a = 5.52 Å, although it is underestimated with respect to the experimental value of a = 5.63 Å [29]. The rhombohedral angle is also comparable to the *LDA+U* value, as well as the relative coordinates of the relaxed rhombohedral structure obtained by the model are in close agreement with the *LDA+U* results. The model spontaneous polarization $P=78$ µC/cm$^2$ is slightly underestimated with respect to the *LDA+U* value of 87.3 µC/cm$^2$ [2]. Concerning the *Pbnm* structure, it is 17 meV/f.u. above the *R3c* ground-state, in excellent agreement with the 14 meV/f.u. obtained with the LDA+U functional. The lattice parameters and internal atomic positions are also in good agreement with the first-principles results.

Figures 1 (a) shows the energy pathway as a function of the atomic displacements that closely connects the two most-stable phases *Pbnm* and *R3c,* left and right panel respectively. All calculations were done in a 40-atom cubic cell with the lattice parameter of the *R3c* structure, *a*=7.80 Å. Note that the energy curves are referred to the *R3c* ground state, and the bridging structure belongs to the *Imma* space group. This phase is characterized by the anti-phase rotation of the oxygen octahedral around the *x* and *y* axes (*a⁻a⁻0* in Glazer's notation), while the Bi displacements are inhibited. Although this phase is not stable (the Bi ions tend to move from their ideal positions), it is useful to reach the phases of interest by simple atomic displacements. The *R3c* structure is obtained by adding anti-phase $O_6$ tilting along [001] (*00a*), plus a polar atomic displacement along [111], whereas the *Pbnm* phase is reached from the *Imma* one with an additional in-phase tilting along [001] *(00a⁺)* and Bi displacements along [110]. The atoms were displaced following the patterns corresponding to each relaxed phase as determined from first-principles calculations. As seen, the model reproduces the BFO instabilities with similar energetic and minimum locations to the *LDA+U* calculations. The *Imma* phase is only 270 meV/f.u. above the *R3c*, which is well below the energy of the prototype cubic phase (870 meV/f.u.). In addition, the model also reproduces the relative phase stability with the volume (Fig. 1 (b)). We obtain a transition pressure between the *R3c* and *Pbnm* phases at 1.5 GPa in agreement with the *LDA+U* value of ~2Gpa [9]. These results indicate that the model description is able to capture the complex energy landscape that characterizes BFO.

## III. RESULTS AND DISCUSSION

We applied the developed interatomic potentials for BFO to simulate the finite temperature behavior of the compound. Molecular Dynamics (MD) simulations were carried out using the DL-POLY code [30] within a constant stress and temperature (N,σ,T) ensemble. In this way, the size and shape of the simulation cell are dynamically adjusted in order to obtain the desired average pressure. A supercell size of 12×12×12 5-

atom unit cells (8640 atoms) was used with periodic boundary conditions. The runs were made at temperature intervals of 50 K, and with a time step of 0.4 fs. Each MD run consists of at least 40000 time steps for data collection after 20000 time steps for thermalization.

### A. Phase diagram

Figure 2 displays the temperature evolution of the relevant magnitudes that characterize the structural behavior of the compound: lattice parameters along pseudo-cubic directions are shown in (a), polar order in (b), and oxygen octahedron tilting patterns in (c). The anti-phase oxygen octahedron tilting is described by

$$\omega^R = \frac{1}{N} \sum_i \omega_i (-1)^{n_x(i)+n_y(i)+n_z(i)}$$

, where $n_x$, $n_y$ and $n_z$ are integers that gives the position of the $i$-cell in the pseudo-cubic box. The in-phase oxygen octahedron tilting along z corresponds to $\omega^M = \frac{1}{N} \sum_i \omega_i (-1)^{n_x(i)+n_y(i)}$ [7]. The polar order is described by the net polarization $P = \frac{1}{N} \sum_i p_i$, where $p_i$ represents the polarization of the perovskite $i$-cell centered at the Bi-cation, and the antiferroelectric order parameter by

$$P^{AF} = \frac{1}{N} \sum_i p_i (-1)^{n_z(i)}$$

which describes the alternating polarization along the [001] direction. The local polarization is defined as the dipole moment per unit volume of a perovskite cell considered here as centered at the Bi atom (A-site), and delimited by the 8 Fe near neighbors at the corners of the box. In the calculations, we take the contributions from all atoms in the conventional cell, and the atomic positions with respect to this center, then $p_i = \frac{1}{v} \sum_i \frac{1}{w_i} z_i (r_i - r_A)$ where $v$ is the volume of the cell, $z_i$ and $r_i$ denote the charge and the position of the $i$-particle respectively, and $w_i$ is a weight factor equal to the number of cells to which the particle belongs. The reference position, $r_A$, corresponds to the core of the Bi atom, and the sum extends over 41 particles, including cores and shells of the surrounding ions (8 Fes and 12 Os) and the shell of the Bi atom. Note that $p$ is independent of the origin for $r_i$ vectors, and it is null when the atoms are at the ideal cubic positions. In the figure, the low-temperature *R3c* phase is characterized by identical values of cell parameters (*a=b=c*), a macroscopic polarization pointing along the [111] direction ($P_x = P_y = P_z$), and an anti-phase octahedral tilting pattern along the [111] direction ($\omega^R_x = \omega^R_y = \omega^R_z$). We note that finite-temperature results are in reasonable agreement with experimental values, even though the underestimation of cell parameters as a result of the LDA input data used to fit the potentials. At room-temperature, for instance, the model gives a spontaneous polarization $P$=77 μC/cm$^2$ which is comparable with the 60-100 μC/cm$^2$ reported for single crystals [31]. The obtained oxygen octahedron tilting angle around the polar axis of 13.8º, agrees well with 11º-13º reported in the experiments [32].

A phase transition with a strong first-order character takes place at $T_C$ = 1100 K, where sudden changes in the system properties are observed. Lattice parameters, that came growing with temperature in the *R3c* phase, display a marked reduction, being more pronounced the change along the *z* direction. The simulated volume contraction ~1.98% is close to the experimental value of 1.56% [33]. In addition, the net polarization $\boldsymbol{P}$ vanishes at $T_C$ ($P_x = P_y = P_z = 0$). It is observed, however, that another type of ordering emerges. It is clear from the phase diagram that the new order corresponds to antipolar $xy$-planes with polarization along [110] directions ($P_x^{AF} = P_y^{AF}$, $P_z^{AF} = 0$). Regarding the oxygen octahedron tilting, it becomes in-phase along z ($\omega^M_z \neq 0$) while it remains out-of-phase along the other two directions ($\omega^R_x = \omega^R_y \neq 0$). We thus conclude that the structure above $T_C$ has *Pbnm* symmetry, in coincidence with neutron diffraction results [33]. The value of $T_C$ is also correctly reproduced by our model description. We note, however, that one might have expected an underestimation of $T_C$ as a result of the *LDA* approximation used to fit the model.

In the simulations, the *Pbnm* phase remains stable over a wide range of temperatures and the system becomes cubic at ~ 2400 K. Above this temperature, the lattice constants are equivalents and the order parameters related to the polarization and the oxygen tilts are negligible. These results are in disagreement with experimental observations where the *Pbnm* phase is stable over nearly 100 K above $T_C$. The system transforms to a cubic metallic phase at 1204 K and melts at 1240 K [34,35]. We note that the shell-model description is not appropriate to describe a metallic phase. The model description assumes that the electronic structure of the compound does not suffer significative changes with the temperature, specifically that the band gap is large enough to avoid any electronic transition. Then, the model is valid to investigate the temperature driven *R3c-Pbnm* phase transition at 1100 K, because both phases are not metallic. However, our description definitively is not longer valid for temperatures over ~1200 K. Instead, the model describes a hypothetical situation where BFO remains always in an ideal non-conducting state. In this regard, our results indicate that the *Pbnm* phase would be stable over a wider temperature range if the metallic behavior is inhibited. They also suggest that the experimentally observed cubic phase is related to a metal-insulator transition rather than to the energy landscape of the non-conducting system obtained from first principle calculations.

We did not find in our simulations evidence of any intermediate phase between the *R3c* and the *Pbnm* structures. A recent study with an effective Hamiltonian approach has proposed the existence of a family of bridging phases with monoclinic symmetry [7]. Those phases are characterized by different octahedral tilting patterns along the z-axis where adjacent octahedron planes can be oriented either in-phase or out of phase, forming quasi-degenerate twinned octahedral tilting structures. We believe that the marked volume reduction at $T_C$, which is correctly reproduced by our model, favors the stabilization of the *Pbnm* phase over other possible metastable structures. Indeed, first-principles calculations show a noticeable volume effect on the phase stability in BFO [9]. This effect is observed in Figure 1 (b), where the energy of the *Pbnm* structure obtained at the *R3c* volume is substantially higher than the one corresponding to the fully relaxed structure.

### B. Pbnm phase characterization

In order to better understand some of the characteristics of the *Pbnm* phase, we follow the description of an antiferroelectric system originally proposed by Kittel [36]. According to this, we decompose the system into two interpenetrating sublattices labeled *a* and *b*. Here, each sublattice includes cells that belong to a given *(001)* plane and to every other plane. Figure 3 shows the local polarization distributions of the sublattices at 1100 K and confirms that both sublattices are identical but with opposite-oriented polarizations along the [110] direction. In each sublattice, the distributions are unimodal with peak positions centered near the average values $p_x = p_y = \pm 14$ µC/cm$^2$ and $p_z = 0$. This antiferroelectric type of ordering is driven by the off-centering of the Bi$^{+3}$ ions, which gives rise to local polarizations, and it is allowed by the symmetry of the phase [37]. The possibility of this type of order in BFO was addressed from first principle calculations. According to that study, the optimized *Pbnm* structure can be aptly described as AFE due to the large antipolar displacements of the Bi cations [9]. While these displacements at zero temperature are well reproduced by our shell model description, the simulations indicate that they do not vanish at temperatures above $T_C$, where the *Pbnm* phase stabilizes.

On the other hand, the *Pbnm* phase has been seen as paraelectric in experiments. Nevertheless, we realize that the structural parameters obtained from neutron diffraction experiments are compatible with an AFE type of distortion. [33] In particular, the Bi displacements in an antipolar fashion along the orthorhombic b-axis (a-axis in *Pnma* and [110] pseudo-cubic direction) can be deduced from the published atomic positions data. Combining these data with our model potential, we estimate that the polarization of the sublattices in the experimental structure would correspond to ~37 μC/cm². This result indicates that an AFE order is indeed present in the real compound in agreement with the model prediction. Furthermore, we would like to stress that the observed AFE order is consistent with the strong tendency of the Bi cations to break the local inversion symmetry due to the stereochemically active lone-pair on Bi cations [38].

The above results are also useful to gain insight on the dynamical behavior of the *R3c*-*Pbnm* phase transition. Ferroelectric transitions have been typically classified as being either of order-disorder or displacive type, and that type can be distinguished by inspecting the distribution of the local polarization just above $T_C$. In the *Pbnm* phase of BFO, the local polarizations can take opposite values (see Figure 3) generating a bimodal distribution. That distribution, however, must not be assigned to an order-disorder type behavior. In an order-disorder dynamics, local polarizations are hopping between equivalent directions in a relaxational motion. This feature, which is observed in the paraelectric phase of perovskites like $BaTiO_3$ or $KNbO_3$, is not present in the case of BFO. The inset of Figure 3 shows the time evolution of the local polarization of one arbitrary unit cell where only the component along the polar direction (pseudocubic [110] direction) is displayed for simplicity. The local polarization shows fast oscillations around a finite value, which is in concordance with the average polarization of the corresponding sublattice, without polarization reversals. Therefore the bimodal distribution of the local polarizations corresponds to the superposition of two ordered sublattices, where individual cells adopt only one polar value according to the sublattice to which they belong. We thus conclude that the *R3c-Pbnm* transition has a displacive-like character, where the atoms move from one equilibrium position to another at $T_C$.

### C. Electric-field induced transition

To gain more insight into the AFE nature of the *Pbnm* phase, we apply an external electric field (E) along the pseudo-cubic [110] direction, which corresponds to the orthorhombic *b* axis. Figure 4 (a) shows the components of the total polarization as a function of E where a double-hysteresis loop, which is typical of an antiferroelectric compound, is observed. In the low field regime, an ordinary dielectric response is observed. The polarization components along the filed direction ($P_x$ and $P_y$) increase linearly as the field increases while the component in the perpendicular direction ($P_z$) is unaffected. When the field reaches the critical value of E = 1.4 MV /cm, a field-induced transition to a ferroelectric phase takes place. At this field, $P_x$ and $P_y$ change abruptly from 8 µC/cm² to 42 µC/cm². Interestingly, the polar component perpendicular to the field $P_z$ is also developed; it changes abruptly from zero to 35 µC/cm². As a result, the net polarization is not collinear with the field direction but lies practically along the pseudo-cubic [111] direction. As the field is further increased, $P_x$ and $P_y$ recover a monotonic behavior while $P_z$ keeps relatively unchanged. The polar structure remains stable as E is released from high values, and the field-induced ferroelectric phase disappears at E = 0.2 MV / cm. The system displays a similar behavior when the field is reversed. We must point out that the theoretical values can not be compared directly with experiments. The changes at the transitions obtained by the simulations are sharper than in experiments, and the theoretical critical field is expected to be larger than in real systems [39].

The field-induced ferroelectric transition is also accompanied by a large volume expansion as shown in Figure 4 (b). This behavior is consistent with observations in AFE compounds, [34,35] and is originated by the difference in volume of the phases below and above the transition. Even more, the behavior of the strain components also supports an AFE-ferroelectric transition rather than a paraelectric-ferroelectric one. If it were the last case, we would expect a different sign in the longitudinal and transversal components of the strain. Nevertheless, the volume expansion obtained for BFO results from the positive contribution of both components of strains. That is, the system expands in both directions, parallel and perpendicular to the electric field during the transition, which is in agreement with the behavior reported in AFE compounds. [35]

To perform a deeper microscopic characterization, Figure 5 (a) shows the P .vs. E curve corresponding to each one of the opposite-polarized sublattices. Here $P_z$ is not shown for simplicity since it evolves in a similar way in both sublattices, i.e. its behavior corresponds to the one displayed in Figure 4(a). Each polar sublattice exhibits a particular hysteresis loop. As it is expected, the polarization of the sublattice aligned antiparallel to the field, $P^a$, is reversed. In addition, its magnitude also increases with respect to the AFE value, changing from -13 μC/cm² to 40 μC/cm². The polarization of the other sublattice, $P^b$, which is already aligned with the field, also increases at the transition. At the end, there is no appreciable difference between the polarizations of the two sublattices in the induced phase, $P^a = P^b$. What is really very interesting is that the applied electric field also produces changes in the oxygen rotation pattern. To see that, it is also convenient to decompose the octahedron tilting in terms of the tilting of the two sublattices $a$ and $b$, labeled as $\omega^a$ and $\omega^b$, and express the rotational order parameters as $\omega^R = \frac{1}{2}(\omega^a - \omega^b)$ and $\omega^M = \frac{1}{2}(\omega^a + \omega^b)$. Figure 5(b) shows the evolution of the z-component of the octahedral tilting of each sublattice as function of the electric field; the x- and y-components (not shown here) do not experience significant variations. We observe that both sublattices display a hysteretic curve, but only one, $\omega^b_z$, changes the sign at the transition. Then, while both sublattices are in-phase ($\omega^a_z = \omega^b_z$) for fields below the transition, they are out-of-phase ($\omega^a_z = -\omega^b_z$) above it. In other words, the rotational order changes from $\omega^M$ to $\omega^R$ as a consequence of the applied electric field. Based on the analysis of the polar and rotational order parameters, we conclude that the field-induced phase recovers the symmetry of the low temperature ferroelectric *R3c* phase. Notably, this electric-field induced transition is similar to the one proposed to occur in rare-earth substituted BFO. [36] Therefore, we also conclude that the AFE nature observed in those solid solutions arises due to an inherent property of the *Pbmn* phase of pure BFO.

The electric-field induced transition shows an unexpected result, the appearance of a polar component perpendicular to the applied electric field with a simultaneous change in the rotational pattern. This behavior reveals the presence of a strong coupling between the rotational and polar degrees of freedom. A simple way to describe the interplay between these order parameters is as follows: the electric field induces the polarization flipping of the sublattice polarized anti-parallel to the field. The switching mainly involves the displacements of the bismuth atoms, which are forced to move from the [-1-10] to the [110] direction. The resulting polar configuration, however, is not compatible with the original oxygen octahedron rotation pattern, and consequently one sublattice has to modify the rotational pattern in order to stabilize the structure. In other words, the Bi ions force the change in the rotational pattern from $a^-a^-c^+$ to $a^-a^-a^-$. The new rotational order allows a stronger polar correlation, and favors the development of a polar component in the direction perpendicular to the electric field. The resulting atomic configuration produces changes in the volume and the shape, and the system goes back to the symmetry of the low temperature phase.

The coexistence of the antipolar atomic distortions with symmetry-allowed oxygen octahedron rotations is a common feature observed in AFE perovskites, and the interplay between them is just starting to be explored [40–42]. In this regard, our results show the key role played by the oxygen octahedron rotations during the electric-field induced transition, and consequently over the properties of interest for technological applications. [39,43].

## IV. CONCLUSIONS

In summary, we demonstrate that the high temperature *Pbnm* phase of pure BFO is anti-ferroelectric by using an atomistic model fitted to first-principles calculations. The description shows a direct transition from the ferroelectric *R3c* phase to the *Pbnm* phase without evidence of any intermediate bridging phase between them. The stabilization of the *Pbnm* phase over other possible structures is attributed to the large volume contrac-tion that takes place at the transition. We show that the *Pbnm* phase displays local polarizations forming an antiferroelectric ordering and a characteristic double-hysteresis loop under an applied electric field. The field-induced transition is also accompanied by large changes of strains. The induced ferroelectric phase recovers the symmetry of the low-temperature *R3c* phase as a result of the strong coupling between the atomic polar displacements, the oxygen octahedron rotations and the strain. We hope that the reported antiferroelectricity in the *Pbnm* phase of BFO, hard to characterize experimentally contributes to a better understanding of the *R3c-Pbnm* morphotropic phase boundary observed in BFO-based solid solutions.

**ACKNOWLEDGMENTS**

We acknowledge computing time at the CCT-Rosario Computational Center. The work was sponsored by Consejo Nacional de Investigaciones Científicas y Tecnológicas (CONICET), Agencia Nacional de Promoción Científica y Tecnológica (ANPCyT), Universidad Nacional de Rosario, and Universidad Nacional del Litoral.

**Table I**: Parameters for the Shell model potential. Units of energy, length and charge are given in eV, Å, and electrons respectively.

| Atom | Core Charge | Shell Charge | $k_2$ | $k_4$ |
|---|---|---|---|---|
| Bi | 6.1856 | -3.2426 | 220.67 | 9900 |
| Fe | 2.3346 | -0.4635 | 495.18 | 0 |
| O | 0.5004 | -2.1051 | 20 | 3000 |
| Short Range | | A | $\rho$ | B |
| Bi-Fe | | 254.15 | 0.4667 | 52.66 |
| Bi-O | | 2863.21 | 0.3028 | 56.83 |
| Fe-O | | 3225.3 | 0.2328 | 28.54 |
| O-O | | 215.64 | 0.495 | -94.66 |
| Three-Body | $k_B$ | | $\theta_0$ | |
| Fe-O-Fe | 2.3808 | | 180 | |

**Table II**: Optimized structures for BiFeO₃ in space groups *R3c* and *Pbnm* obtained with the model and LDA calculations. Experimental values are include for comparison

| Atom | Wyc. | Position | Model | LDA+U | Exp. [29] |
|---|---|---|---|---|---|
| | | | R3c | | |
| Bi | 2a | x | 0 | 0 | 0 |
| Fe | 2a | x | 0.221 | 0.227 | 0.221 |
| O | 6b | x | 0.531 | 0.54 | 0.538 |
| | | y | 0.935 | 0.94 | 0.933 |
| | | z | 0.394 | 0.40 | 0.395 |
| | a = b = c = | | 5.52 Å | 5.52 Å | 5.63 Å |
| | α = β = γ = | | 59.1° | 59.84° | 59.35° |

| Atom | Wyc. | Position | Model | LDA+U | Exp. |
|---|---|---|---|---|---|
| | | | Pbnm | | |
| Bi | 4c | x | -0.002 | 0.009 | -0.001 |
| | | y | 0.038 | 0.052 | 0.023 |
| | | z | 0.25 | 0.25 | 0.25 |
| Fe | 4b | x | 0.5 | 0.5 | 0.5 |
| | | y | 0 | 0 | 0 |
| | | z | 0 | 0 | 0 |
| O | 4c | x | 0.595 | 0.594 | 0.566 |
| | | y | 0.972 | 0.972 | 0.979 |
| | | z | 0.25 | 0.25 | 0.25 |
| O | 8d | x | 0.304 | 0.303 | 0.212 |
| | | y | 0.3 | 0.299 | 0.289 |
| | | z | 0.046 | 0.046 | 0.032 |
| | a | | 5.42 Å | 5.37 Å | 5.613 Å |
| | b | | 5.54 Å | 5.59 Å | 5.647 Å |
| | c | | 7.66 Å | 7.69 Å | 7.971 Å |

**Figure Captions**:

**Figure 1**: (Color online) Total energy at different configuration obtained with the model (solid lines) and with *LDA+U* calculations (dash lines). (a) Energy as functions of atomic displacements from the *Imma* phase towards the *Pbnm* (left) and *R3c* (right) ones. Atoms are moved following the pattern of the *LDA* relaxed structure, which is taken as the unit. All calculations were done in a 40-atom cubic cell with *a*=7.80 Å. (b) Energy as function of the volume for relaxed *Pbnm* and *R3c* structures.

**Figure 2**: (Color online) Phase diagram of BiFeO$_3$ at zero pressure resulting from the MD simulations. Thermal evolution of the lattice parameters (a), net polarization, *P*, and order parameter associated with the antiferroelectric order along the *z*-direction, *P*$^{AF}$ (b), and the oxygen octahedral rotation parameters describing anti-phase, $\omega^R$, and in-phase $\omega^M$, tilting along the *z*-direction(c).

**Figure 3:** (Color online) Distributions of local polarizations corresponding to the *Pbnm* phase at 1100 K. The superscripts correspond to the sublattices in which the system can be decomposed. Inset: Time evolution of one unit-cell polarization along the [110] direction.

**Figure 4:** (Color online) Polarization (a) and strains (b) as function of the electric field obtained from the *Pbnm* phase at 1100 K. The field is applied along [110] directions, and arrows indicate field changes.

**Figure 5:** (Color online) Decomposed sublattice polarizations (a) and oxygen octahedral rotation (b) as function of the applied electric field.

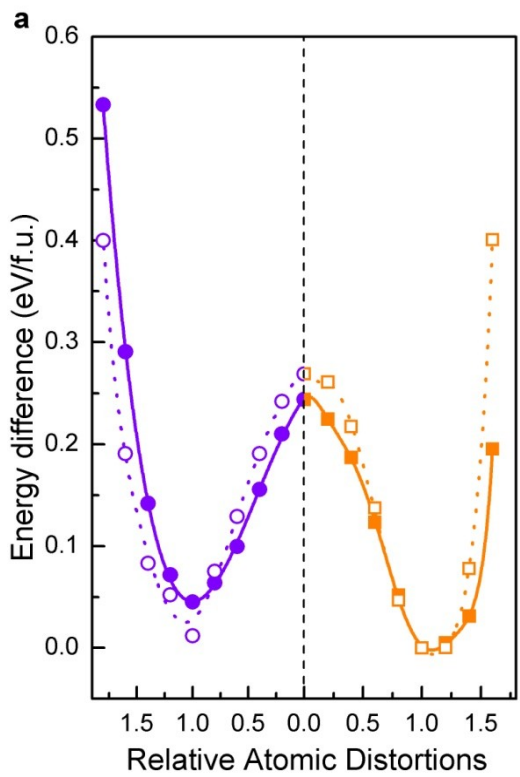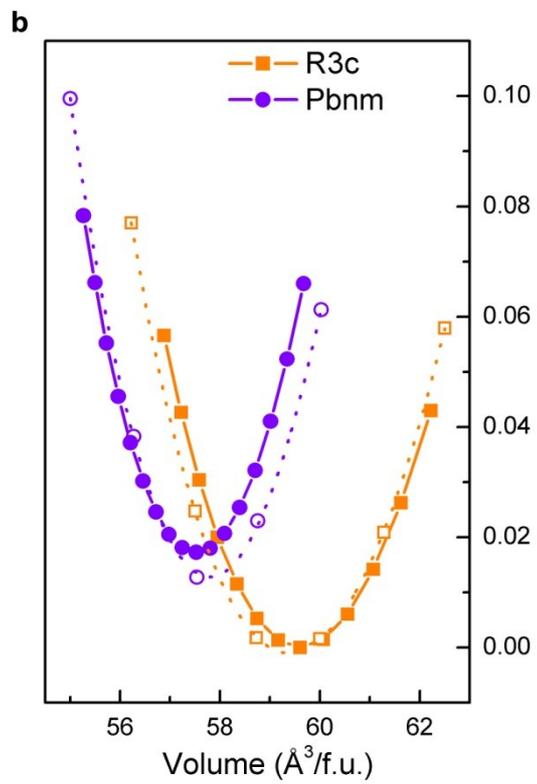

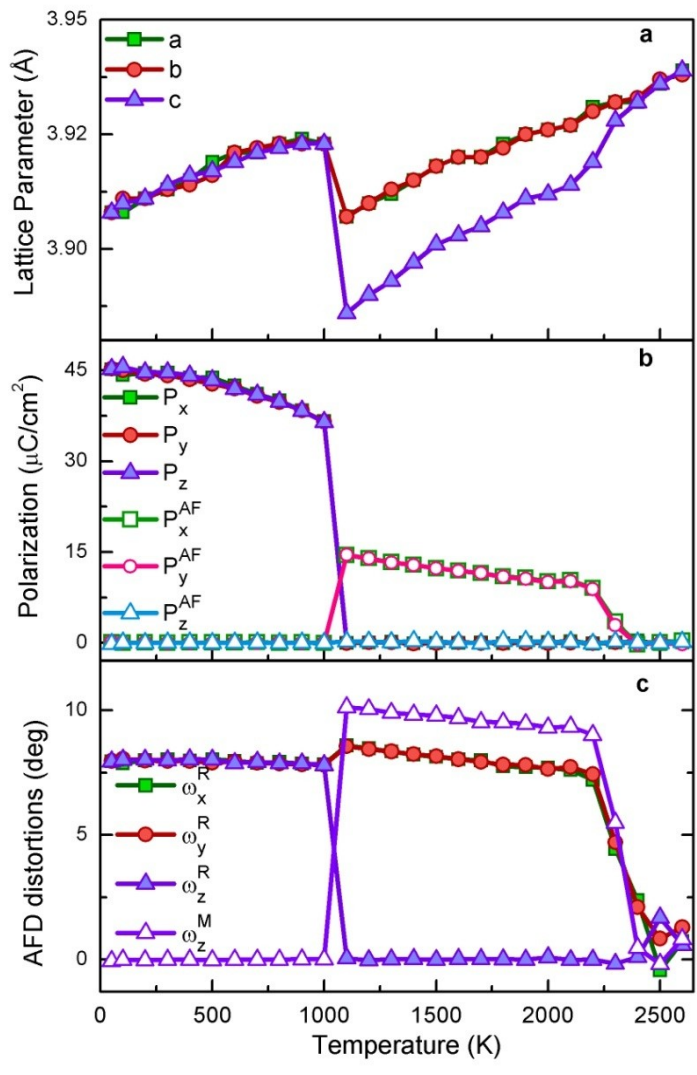

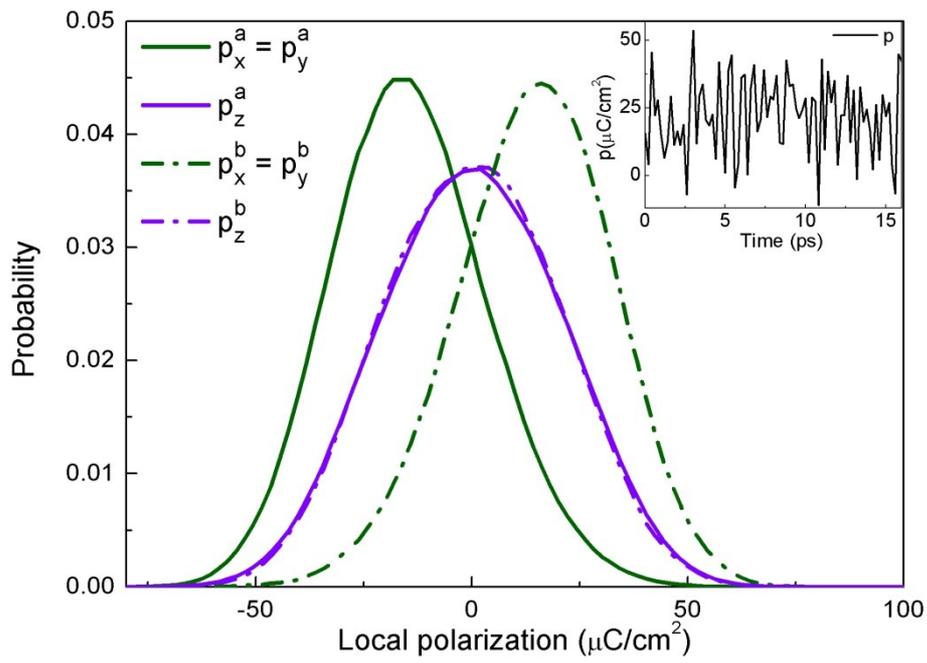

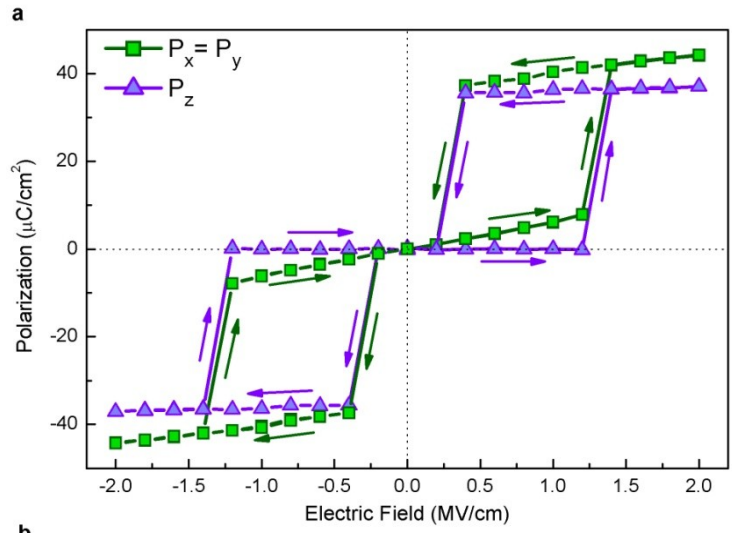

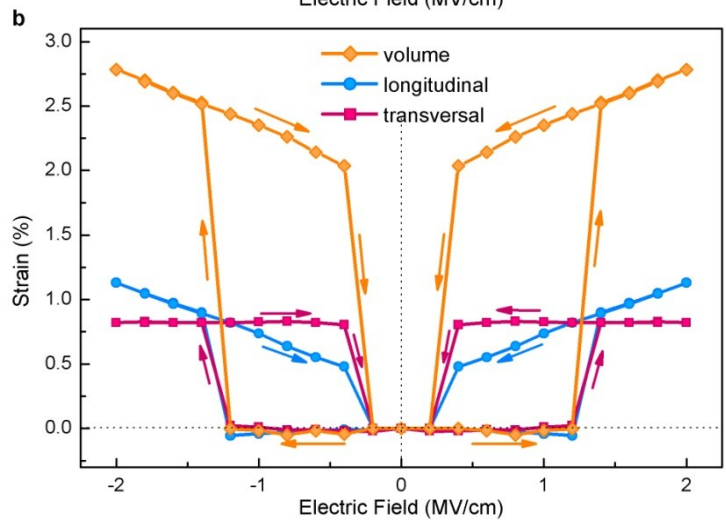

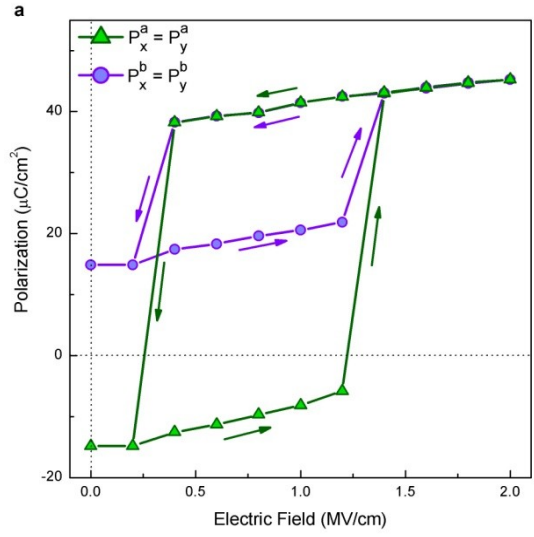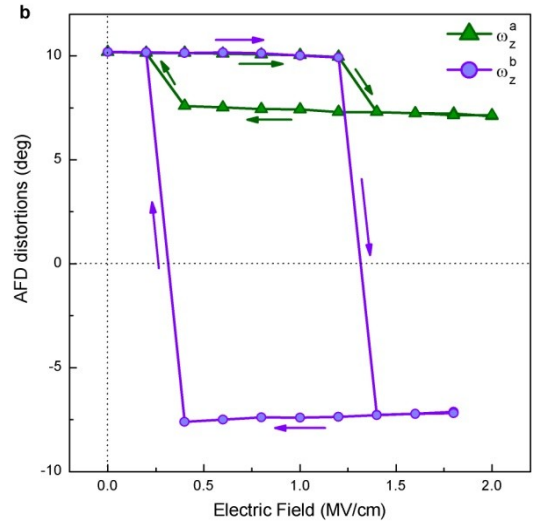